# Response to "Feasibility of 3D reconstruction from a single 2D diffraction measurement"

Jianwei Miao

Department of Physics and Astronomy and California NanoSystems Institute,
University of California, Los Angeles, CA 90095, USA.
Email: miao@physics.ucla.edu.
(Dated: 9/17/2009)

ABSTRACT

We present our response to Pierre Thibault's article, titled "Feasibility of 3D reconstruction from a single 2D diffraction measurement" [1], in which he commented upon our recent ankylography paper [2]. While we appreciate Thibault's efforts in promoting further reflection on our paper, we found serious errors both in his understanding and analysis of ankylography: he inaccurately presented the oversampling scheme in ankylography, incorrectly described our reconstruction algorithm and our experiment, and formulated arguments based upon a flawed and overly-restrictive theoretical analysis. Therefore, we conclude that his main claims are either scientifically invalid or a misrepresentation of our claims about ankylography. Below is our detailed point-by-point response to his criticisms.

## 1. Response to "Section II. A. Preliminaries"

In this section, Thibault implied the requirement of obtaining the autocorrelation function for phase retrieval of coherent diffraction intensities, which was the basis of his theoretical analysis. This is, unfortunately, inaccurate. Historically, Sayre first realized the oversampling concept for potential phase retrieval [3], which was based upon the fact the autocorrelation function of an object is exactly twice of the size of the object. Later, it was mathematically shown that, when the diffraction pattern of a 2D or 3D object is sampled at a frequency at least twice finer than the Nyquist interval, the phases are in principle uniquely encoded inside the diffraction pattern (defined as statement 1) [4-7]. Note that in some rare cases, the phases are not unique, no matter how fine the diffraction intensities are sampled [8]. In 1998, based on numerical experiments of both noise-free and noisy data, Miao, Sayre and Chapman showed that it is *unnecessary* to sample the diffraction pattern by twice in each dimension for successful phase retrieval of 2D and 3D objects [9]. They proposed that, when the number of measured independent intensity points is more than the number of unknown variables, the phases can usually be retrieved from the diffraction intensities (defined as statement 2), which was further confirmed by experimental results [10]. Since then, a number of groups has verified this statement by using numerical experiments, as can be easily done. Although a rigorous mathematical proof of this statement is still missing, statement 2 is more general and fundamental than statement 1, as the former is consistent with both the numerical and real experimental results [9,10] and the latter is overly restrictive. One may argue that while statement 2 is true for noise-free or low noise data, what if there is high noise in the data? Certainly it is very difficult, if not possible, to include noise in a mathematical model. As a matter of fact, even the theoretical papers dealing with statement 1 did not include a noise analysis [5,7]. We hence conclude that *it is unnecessary to know the autocorrelation function for successful phase retrieval of 2D and 3D objects*. Unfortunately, retrieving the autocorrelation function from the diffraction intensities is the main theme in Thibault's theoretical analysis in ref. 1.

## 2. Response to "Section II. B Scattering in the Born approximation"

In this section, Thibault raised a question about digital in-line holography vs. ankylography. During the period of forming the ankylography concept, we reviewed the literature on holography, including such papers presented in the book "Three-Dimensional Holographic Imaging" [11]. From this, we concluded that there exits several fundamental differences between holography and ankylography. First of all, in-line holography mainly deals with near field, while our ankylography paper deals with far field. Second, holography requires a reference wave and its resolution is limited by either the detector resolution or the source size of the reference wave. On the other hand, ankylography is based upon coherent diffraction and does not require a reference wave. The resolution of ankylography is only determined by the diffraction angle (*i.e*. Eq. 2 in ref. 2). Finally, although a hologram encodes some depth information, holography can not generally recover true 3D structure from a single hologram (as Thibault also points out). But we demonstrated that, through oversampling the spherical diffraction intensities and enforcing the physical constraints, ankylography can reconstruct the 3D structure from a single view [2].

## 3. Response to "Section II. C Non-uniform sampling"

This is the key section for Thibault's theoretical analysis. Unfortunately, his presentation of the oversampling scheme in ankylography is inaccurate, and his theoretical analysis is flawed.

First, Thibault's illustration of the oversampling scheme (Figs. 1e and 2a in ref. 1) will likely confuse general readers (including us), we hence present a clearer version. Fig. 1 (below) shows the oversampling scheme in ankylography, where the array size in Fourier space is N x N pixels and the support size in real space is M x M pixels. The two curves in Fig. 1(a) represent the Ewald sphere and its centro-symmetry (*i.e*. the Friedel pairs). In ankylography the distance between two neighbouring sampling points on the Ewald sphere (*i.e.* red circles) is the same as the distance between two neighbouring grid points in the N x N array (*i.e.* black dots).

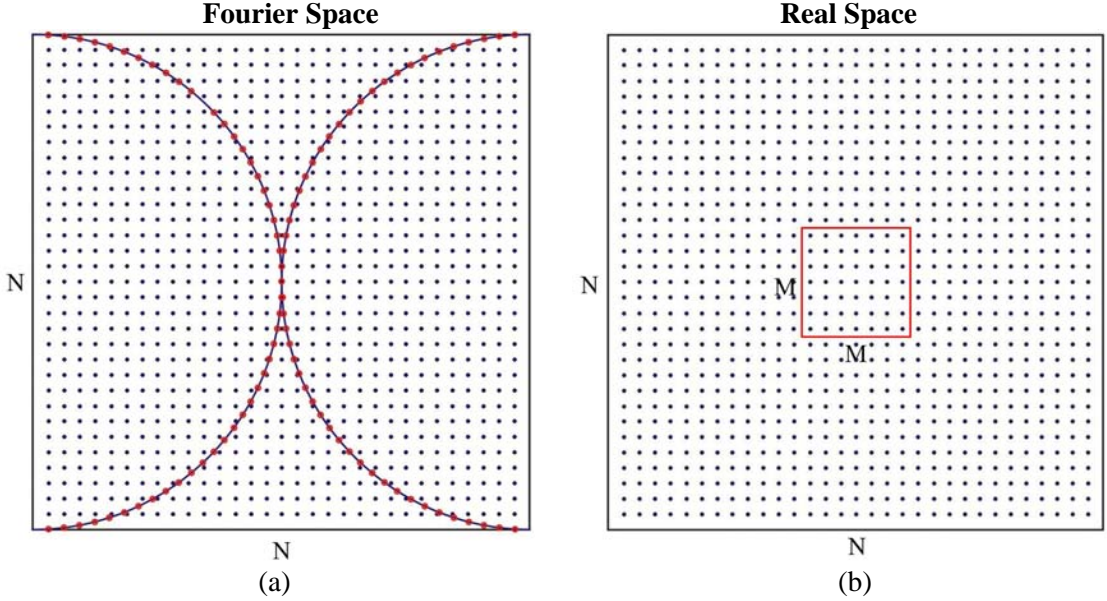

**Fig. 1** The oversampling scheme in ankylography. **(a)** The two half-circles represent the Ewald sphere and its centro-symmetry (*i.e.* the Friedel pairs). The red dots are the intensity points measured on the Ewald sphere. Note that the distance between two neighbouring red dots is the same as the distance between two neighbouring grid points (*i.e.* black dots). **(b)** The red square is the support in real space. The oversampling degree ($O_d$) is defined as the number of red dots on one of the half circles divided by the number of grid points inside the support. In this case, N = 32 and the diffraction angle ($2\theta_{max}$) = 90°.

In our ankylography paper [2], the oversampling degree ($O_d$) was defined as the number of measured points on the Ewald sphere divided by the number of unknown variables inside the support. In a special 2D case, $O_d$ can be written as

$$O_d = \frac{\pi N}{2M^2} \qquad (1)$$

Note that Eq. (1) would be different in 3D cases. Based on the requirement of $O_d > 1$ in ankylography, we obtain that M ~ 7 for the case illustrated in Fig. 1 where N = 32 and $2\theta_{max}$ = 90°. In other words, the number of unknown variables in this case has to be smaller than ~ 49.

Here we want to further clarify the principle of ankylography. When the diffraction pattern is oversampled on a spherical surface as described in Fig. 1, we showed that *the intensity points encode information from all possible orientations of a 3D object* (supplementary information in ref. 2). When the number of measured independent points is more than the number of unknown variables, we argue that the 3D object can usually be recovered from the 2D spherical diffraction pattern alone. We don't have a rigorous mathematical proof yet (as it is a *non-linear* problem), but we intend to demonstrate this statement by using many numerical experiments. In our numerical experiments, we discovered that the distribution of the diffraction pattern on a spherical surface is critical for unique 3D reconstructions. For example, if we used the diffraction intensities on a plane or a few intersected planes, the 3D reconstructions would fail even with a large oversampling degree. The reason is because the diffraction intensities on the intersected planes don't encode information from all possible orientations of the object. By using numerical simulations and an experiment, we suggested in ref. 2 that a general 3D finite object can in principle be reconstructed from a single spherical diffraction pattern alone.

Second, we note that some of the representation of the oversampling scheme shown in Fig. 1e and Fig. 2a in ref. 1 is inaccurate. The correct way to illustrate the figures is shown in Fig. 2 (below), that is, the spatial resolution along the Z-axis (*i.e.* the horizontal axis in Fig. 2) is determined by the height of the spherical cap [see Eq. 2 in ref. 2]. Thibault's representation along the Z-axis in Fig. 1e and Fig. 2a in ref. 1 is related to the super-resolution scheme, but not ankylography.

Third, 3D ankylographic reconstruction is intrinsically a *non-linear* problem. But Thibault treated it as a linear one by attempting to retrieve the autocorrelation function from non-uniformly sampled diffraction intensities. As discussed above in response 1, retrieving the autocorrelation function is *not* a pre-requirement for successful phase retrieval. As a matter of fact, when the diffraction intensities are sampled at a frequency less than twice finer than the Nyquist interval, the autocorrelation is wrapped around and can't be retrieved from the diffraction intensities directly (that is, without first performing phase retrieval). But as long as the number of measured intensity points is more than the number of unknown variables, the phases can still be successfully retrieved.

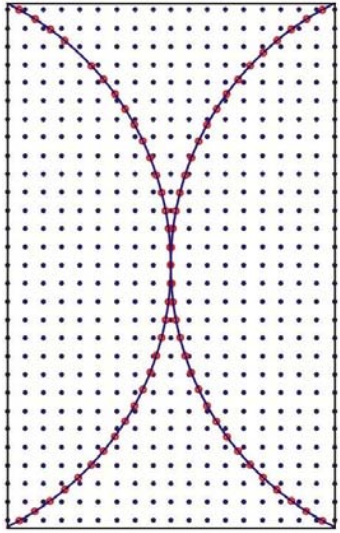

**Fig. 2** When the diffraction angle ($2\theta_{max}$) is smaller than 90°, the intensity points are measured on a spherical cap. The height of the cap determines the depth resolution, that is, the array size of the regular grid points has to be just large enough to include the spherical cap pairs. This figure is to show that Figs. 1e and 2a in ref. 1 are not accurately illustrated.

Finally, compared to coherent diffraction microscopy (or coherent diffractive imaging), the phase retrieval problem in ankylography is in principle more difficult to deal with. The reason is related to the facts that the sampling matrix is ill-posed and the distribution of singular values decays fast (as Thibault pointed out in ref. 1). But, unfortunately, Thibault ignored all the physical constraints in his theoretical analysis, which include positivity, uniformity outside the support, continuity inside the support and amplitude extension [2]. It is well-known in the coherent imaging community that without the physical constraints, it would be very difficult to perform phase retrieval successfully [12]. Additionally, significant research has taken place recently in the applied mathematics community with regards to "compressed sensing" [13,14]. Under certain conditions, a function can be completely determined from a very sparse set of measurements in Fourier space by using the $l_1$ norm constraint (*i.e.* minimization of the sum of the absolute value of the function). While we have reservations about the general applicability of compressed sensing as it only uses the $l_1$ norm constraint, we believe that ankylography is in a much more favourable condition. We not only measure more independent intensity points than the unknown variables, but also enforce many physical constraints. Besides those constraints listed above, one may also apply constraints such as atomicity, histogram matching, molecular replacement and non-crystallographic symmetry in ankylography. Hence we conclude that the most critical flaw in Thibault's theoretical analysis is having ignored the vital importance of the various physical constraints.

**4. Response to "Section III. Reconstruction Algorithm"**

Our ankylographic reconstruction algorithm combines the hybrid-input-output (HIO) algorithm [15] with a number of new constraints, including selection of the best random initial phase set, uniformity outside the support, continuity inside the support and amplitude extension. We believe that enforcing more physical constraints is probably the most useful strategy to improve the reconstruction algorithm. Below is our detailed explanation of the new constraints.

(i) Selection of the best random initial phase set. We usually start with 50 independent reconstructions using random initial phase sets and an updated support (for the first iteration, a loose support is used which can be estimated from the speckle size). We determine the best random initial phase set at the end of 1000 HIO iterations, which corresponds to the minimum $R_{Ewald}$ defined in the supplementary information [2]. We found this step is effective as an optimized random initial phase set can usually result in a good final reconstruction.

(ii) Uniformity outside the support. For noise-free reconstructions, it may be true that the electron density outside the support constraint is zeros. But for the reconstructions of noisy data, those who use the HIO algorithm know that the electron density inside the support sometimes oscillates significantly between iterations, which is due to the noise present in the diffraction intensities. Experimentally we found that, by incorporating the uniformity constraint outside the support in the HIO algorithm, we can reduce the electron density oscillation and hence improve the quality of the reconstructions. A few members in our group have independently tested this constraint and found it very effective.

(iii) Continuity inside the support. In ankylographic reconstructions, the electron density of the reconstructed images should be continuous and there should exist no sharp edges. This is because in coherent diffraction microscopy the amplitude transfer function is convolved with the true wave function of the object. We found it is a useful physical constraint but not as effective as the uniformity constraint.

(iv) Amplitude extension. This is similar to phase extension in X-ray crystallography [16]. But unlike phase extension, we use amplitude extension to retrieve the missing Fourier modulus. We found that, for the reconstructions of a small-array object, we don't need this constraint. But when the object gets larger, the amplitude extension constraint is very useful.

Our ankylographic reconstruction algorithm is discussed in detail in the supplementary information in ref. 2. Many of Thibault's questions are actually addressed in the latest version of our paper [2]. For example, we only selected the best random initial phase set at the beginning of each reconstruction. We applied the uniformity constraint once every 10 iterations. We only applied the continuity constraint once in step 5 in the algorithm described in page 8 of the supplementary information in ref. 2. The implementation of amplitude extension can be found in Tab. S1 [2]. The convergence of our algorithm is monitored by two R-factors, $R_{Ewald}$ and $R_{Entire}$ defined in the supplementary information in ref. 2. As for the definition of a "solution", we first try to make the R-factors as small as possible in the reconstructions (Tabs. S1 and S2 in ref. 2). We then performed independent reconstructions and made sure the results are consistent. For the numerical experiments, we also used the original structure to quantify the reconstructions.

In this section, Thibault also claimed that in the optimization field, a "projection formalism" is preferred which has an advantage over other algorithm structures. However, this is not true. Although, the projection formalism is another classical method for solving optimization problems, it suffers from such shortcomings as slower convergence and tendency to be trapped in local minima. Therefore, people are always interested in more sophisticated iterative scheme other than the projection method. For example, in the phase retrieval problem, the error reduction method is a typical projection method [15], while the HIO method is not. The HIO method employs a more complicated iterative scheme and outperforms the error reduction method considerably [12]. In our ankylographic reconstruction algorithm, many constraints are imposed in a loose manner (instead of a hard projection). We believe this is a natural and appropriate choice, as verified by the results from our numerical experiments [2]. Furthermore, other groups have also shown that by enforcing the constraints in a loose or periodic manner, better reconstructions are attained [17].

Finally, we acknowledge that this is naturally an initial algorithm and that future development may be necessary. As ankylography is such a new idea and potentially an important direction, we certainly can't solve all the problems in this first paper. In order to establish this new direction, as with other fields, a series of papers are needed. In terms of the algorithm development, we or others will still have to figure out how to more effectively incorporate all the physical constraints into the iterative algorithm.

**5. Response to "Section IV. A Flawed Experimental Demonstration"**

We found that Thibault's interpretation of our experiment is incorrect. First of all, his theoretical analysis in this section is self-contradictory. He contributed our experiment to a consequence of a three-dimensional version of Babinet's principle. Specifically, he used the equation $\rho_{mask} = \rho_{slab} - \rho_{figure}$ and then took the Fourier transform on both sides of the equation. He may not have realized that taking the Fourier transform on both sides requires two important assumptions: the Born approximation and the Fraunhofer approximation. After obtaining $\rho_{figure}$, he then pointed out that the Born approximation is invalid because the absorption of $\rho_{figure}$ is very large. But it was incorrect to take the Fourier transform on both sides of the equation in the first place. Secondly, while Babinet's principle holds for the Fraunhofer diffraction intensities from complementary 2D objects [18,19], it is not rigorously valid for 3D objects when the diffraction intensities sampled on a spherical surface. Thus one can't simply apply Babinet's principle to our 3D case. Third, because of incorrectly applying the three-dimensional

Babinet's principle, Thibault's numerical simulations presented in Fig. 3b in ref. 1 are flawed. This can be illustrated by using a thought experiment. Assume that we place an aperture into a laser or coherent X-ray beam where the substrate of the aperture is almost opaque to the beam (in our experiment the transmittance of the sample subtract is about $3.2 \times 10^{-4}$). The diffraction intensities from the aperture in the far field form a Fraunhofer diffraction pattern, when the diffraction angle is small. However, Fig. 3b in ref. 1 doesn't look like a Fraunhofer diffraction pattern at all.

Here we present our interpretation of the experiment. Fig. 3 shows the geometrical construction of our demonstration experiment, where the detector is spherically curved. Although we used a planar CCD detector to measure the diffraction intensities in the experiment, we eventually projected the planar diffraction pattern onto the spherical surface for ankylographic reconstructions (see supplementary information in ref. 2). According to the Huygens-Fresnel principle [20], all the points within the 3D confined volume (*i.e.* our 3D test sample) will generate secondary waves and form an interference pattern on the detector. For example, the waves from $P_1$ and $P_2$ will interfere on the detector, and the phase is determined by the optical path difference. At angles away from the optical axis, the interference pattern encodes the depth information of the sample. One may also notice that for the points very close to the sample edge (*e.g.* $P_3$), some of the waves from those points will be blocked by the edge (*i.e.* the edge effect), which potentially cause a problem for our reconstructions. Fortunately, the size of our test sample is much larger than its thickness and hence this effect is negligible. Based on the above analysis, it can be easily shown by using the diffraction theory [19,20] that, when the edge effect is negligible, the measured intensities are proportional to the square of the Fourier transform of the 3D sample on the Ewald sphere.

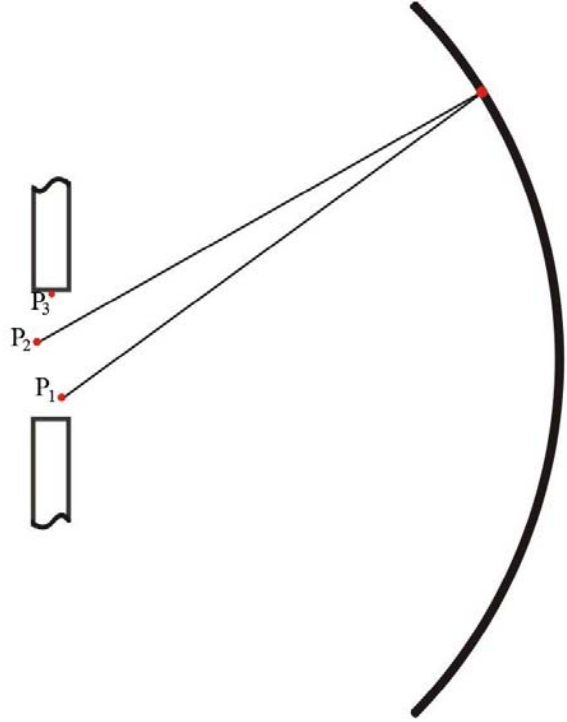

**Fig. 3** Geometrical construction of our experiment where the dimensions are not to scale. According to the Huygens-Fresnel principle, points within the 3D confined volume (such as $P_1$ and $P_2$) will generate secondary waves and form an interference pattern on the detector. At angles away from the optical axis, the interference pattern encodes the depth information. Some of the waves from those points very close to the edge such as $P_3$ are blocked by the edge. However, since the size of our test sample is much larger than its thickness, this effect is negligible.

In this experiment, the test sample was slanted relative to the incident beam, which provides 3D depth that is considerably larger than our resolution. In the 3D reconstruction, the sample information along the Z-axis and the title angle are clearly resolved (Figs. 4b, c and d in ref. 2). Furthermore, we identified two structure defects in the SEM image of the test object (arrows in Fig. 3e in ref. 2), which are spatially resolved in the 3D reconstructed image (arrows in Fig. 3b in ref. 2). These structure defects would be invisible in the 2D reconstructed image. Thus we feel that the combination of the experimental result and our qualitative argument with numerical simulation results convincingly and comprehensively demonstrates both the conceptual soundness and the experimental feasibility of ankylography. This is a new concept, but we have already taken it beyond the "cute idea, but can it ever really work?" stage.

**6. Final Statement**

As ankylography is a new and surprising concept, we enthusiastically welcome any scientific debates about its limitations and potential. We thank Thibault for his "open review" as we feel that it has allowed us to further clarify the intricacies of our ankylography technique. However, we do not agree with his criticisms of our ankylography paper and we stand by the claims therein absolutely. We are happy to provide all the detailed procedures, data sets and the ankylographic reconstruction codes to anyone interested in this new imaging technique.

**Acknowledgements**

We thank the co-authors of the original ankylography manuscript who contributed greatly to this response, especially Kevin S. Raines, Richard L. Sandberg, Huaidong Jiang, Benjamin P. Fahimian and Sara Salha. We also thank David Mao for stimulating discussions on our response to the non-uniform sampling and the reconstruction algorithm sections.